\documentclass[a4paper]{jpconf}
\usepackage{graphicx}
\usepackage{enumitem}
\begin{document}
\title{FAMU: study of the energy dependent \\ transfer rate  $\Lambda_{\mu p \rightarrow \mu O}$ } 

\author{E~Mocchiutti$^{1}$, V~Bonvicini$^{1}$,
  M~Danailov$^{1,2}$, E~Furlanetto$^{1,3}$,
  K~S~Gadedjisso-Tossou$^{1,4,5}$,  D~Guffanti$^{1,6}$,
  C~Pizzolotto$^{1}$, A~Rachevski$^{1}$, L~Stoychev$^{1,4}$,
  E~Vallazza$^{1}$, G~Zampa$^{1}$, J~Niemela$^{4}$, K~Ishida$^{7}$,
  A~Adamczak$^{8}$, G~Baccolo$^{9,10}$, R~Benocci$^{9,11}$,
  R~Bertoni$^{9}$, M~Bonesini$^{9}$, F~Chignoli$^{9}$,
  M~Clemenza$^{9,10}$, A~Curioni$^{9}$, V~Maggi$^{9,11}$,
  R~Mazza$^{10}$, M~Moretti$^{9,10}$, M~Nastasi$^{9,10}$,
  E~Previtali$^{9}$, D~Bakalov$^{12}$, P~Danev$^{12}$,
  M~Stoilov$^{12}$, G~Baldazzi$^{13,14}$, R~Campana$^{13}$,
  I~D'Antone$^{13}$, M~Furini$^{13}$, F~Fuschino$^{13,15}$,
  C~Labanti$^{13,15}$, A~Margotti$^{13}$, S~Meneghini$^{13}$,
  G~Morgante$^{13,15}$, L~P~Rignanese$^{13,14}$, P~L~Rossi$^{13}$,
  M~Zuffa$^{13}$, T~Cervi$^{16,17}$, A~De~Bari$^{16,17}$,
  A~Menegolli$^{16,17}$, C~De~Vecchi$^{17}$, R~Nard\`o$^{17}$,
  M~Rossella$^{17}$, A~Tomaselli$^{17,18}$, L~Colace$^{19,20}$,
  M~De~Vincenzi$^{19,21}$, A~Iaciofano$^{19}$, F~Somma$^{19,22}$,
  L~Tortora$^{19}$, R~Ramponi$^{23}$, and A~Vacchi$^{1,3,7}$}

\address{$^1$ National Institute for Nuclear Physics (INFN), Sezione
  di Trieste, via A. Valerio 2, 34127 Trieste, Italy}
\address{$^2$ Elettra-Sincrotrone Trieste S.C.p.A., SS14, Km 163.5,
  34149 Basovizza, Trieste, Italy}
\address{$^{3}$ Mathematics and Informatics Department, Udine University, via delle Scienze 206, Udine, Italy}
\address{$^4$ The Abdus Salam International Centre for Theoretical
  Physics, Strada Costiera 11, Trieste, Italy}
\address{$^5$ Laboratoire de Physique des Composants \`a
  Semi-conducteurs (LPCS), D\'epartement de physique, Universit\'e de Lom\'e, 01 BP 1515 Lom\'e, Togo}
\address{$^6$ Gran Sasso Science Institute - INFN-LNGS, via F. Crispi
  7, L'Aquila, Italy}
\address{$^7$ RIKEN-RAL RIKEN Nishina Center for Accelerator-Based Science, 2-1, Hirosawa, Wako, Saitama 351-0198, Japan}
\address{$^8$ Institute of Nuclear Physics, Polish Academy of
  Sciences, Radzikowskiego 152, PL31342 Krak\'{o}w, Poland}
\address{$^9$ National Institute for Nuclear Physics (INFN), Sezione
  di Milano Bicocca, piazza della Scienza 3, Milano, Italy}
\address{$^10$ Universit\`a di Milano Bicocca, Dip. di Fisica ``G. Occhialini'', piazza della Scienza 3, Milano, Italy}
\address{$^{11}$ Universit\`a di Milano Bicocca, Dip. di Scienze dell'Ambiente e della Terra, piazza della Scienza 1, Miano, Italy}
\address{$^{12}$ Institute for Nuclear Research and Nuclear Energy,
  Bulgarian Academy of Sciences, blvd. Tsarigradsko ch. 72, Sofia 1142, Bulgaria}
\address{$^{13}$ National Institute for Nuclear Physics (INFN),
  Sezione di Bologna, viale Berti Pichat, 6/2, Bologna, Italy}
\address{$^{14}$ Department of Physics and Astronomy, University of Bologna, viale Berti Pichat, 6/2, Bologna, Italy}
\address{$^{15}$ INAF-IAFS Bologna, Area della Ricerca, via P. Gobetti 101, Bologna, Italy}
\address{$^{16}$ Department of Physics, University of Pavia, via Bassi
  6, Pavia, Italy}
\address{$^{17}$ National Institute for Nuclear Physics (INFN),
  Sezione di Pavia, via Bassi 6, Pavia, Italy}
\address{$^{18}$ Department of Electrical, Computer, and Biomedical
  Engineering, University of Pavia, via Ferrata 5, Pavia, Italy}
\address{$^{19}$ National Institute for Nuclear Physics (INFN),
  Sezione di Roma Tre, via della Vasca Navale 84, Roma, Italy}
\address{$^{20}$ Dipartimento di Ingegneria Universit\`a degli Studi Roma Tre, via V. Volterra, 62, Roma, Italy}
\address{$^{21}$ Dipartimento di Matematica e Fisica, Universit\`a di
  Roma Tre, via della Vasca Navale 84, Roma, Italy}
\address{$^{22}$ Dipartimento di Scienze, Universit\`a di Roma Tre,
  viale G. Marconi 446, Roma, Italy}
\address{$^{23}$ IFN-CNR, Department of Physics - Politecnico di
  Milano and National Institute for Nuclear Physics (INFN), Sezione di
  Milano Politecnico, piazza Leonardo da Vinci 32, 20133 Milano, Italy}

\ead{Emiliano.Mocchiutti@ts.infn.it}

\begin{abstract} 
The main goal of the FAMU experiment is the measurement of the
hyperfine splitting (hfs) in the 1S state of muonic hydrogen $\Delta
E_{hfs}(\mu^-p)1S$. The physical process behind this experiment is the following: $\mu p$ are formed in a mixture of
hydrogen and a higher-Z gas. When absorbing a photon at resonance-energy $\Delta E_{hfs}\approx0.182$~eV, in subsequent collisions with the
surrounding $H_2$ molecules, the $\mu p$ is quickly de-excited and accelerated by $\sim2/3$ of the excitation energy. The observable is the
time distribution of the K-lines X-rays emitted from the $\mu Z$ formed by
muon transfer $(\mu p) +Z \rightarrow (\mu Z)^*+p$, a reaction whose
rate depends on the $\mu p$
kinetic energy. The maximal response, to the tuned laser wavelength,
of the time distribution of X-ray from K-lines of the $(\mu Z)^*$ cascade
indicate the resonance.
During the preparatory phase of the FAMU experiment, several
measurements have been performed both to validate the methodology and
to prepare the best configuration of target and detectors for
the spectroscopic measurement. We present
here the crucial study of the energy dependence of the transfer rate
from muonic hydrogen to oxygen ($\Lambda_{\mu p \rightarrow \mu O}$),
precisely measured for the first time.  
\end{abstract}

\section{Introduction}
\label{intro}
FAMU (Fisica degli Atomi Muonici) is a high precision spectroscopy experiment, conceived 
to exploit exotic atoms properties to study and test the quantum
electro-dynamics (QED).

The main goal of the FAMU experiment is to measure for the first time the hyperfine splitting of the $\mu p$ ground 
state.  Through the measurement of the hyperfine splitting of the $\mu p$ ground 
state it is possible to determine the proton 
Zemach radius. In literature, the ``standard'' measurement of the Zemach
radius of the proton $R_p$ is achieved using ordinary hydrogen. A comparison with the value extracted from muonic hydrogen 
may reinforce or delimit the proton radius puzzle \cite{pohl10}.

Muonic hydrogen atoms ($\mu p$) are formed in a 
hydrogen gas target. In subsequent collisions with $H_2$ molecules, the
$\mu p$ de-excite to the thermalized $\mu p$ in the ground state
($(1S)_{F=0}$). At this point, a variable frequency mid infrared laser light is used to illuminate the gas
target. When the frequency is tuned on the hyperfine splitting
resonance, singlet-to-triplet transitions are induced. A spontaneous radiative  de-excitation of the $(1S)_{F=1}$ state of $\mu
p$ atoms has a very low probability and thus collisions with neighboring molecules 
 establish an effective mechanism of spin-flip de-excitation~\cite{gershtein}.
These collisions are non-radiative processes so that photons are not
observed. Due to momentum conservation, the transition energy is
partially converted to kinetic energy
that sum up to the thermal kinetic energy of the $\mu p$ system. The
$\mu p$ atom in this status is called ``epithermal'' $\mu p$. The
average kinetic energy acquired by the $\mu p$ is about two-thirds of the hyperfine 
transition energy ($\approx 120$~meV).

In the FAMU experiment the energy dependence of the muon 
transfer from muonic hydrogen to another higher-Z gas is exploited to 
detect the occurred transition in $\mu p$. In general, the
muon-transfer rate at low energies $\Lambda_{\mu p \rightarrow \mu Z}$ is energy
independent. However, for a few gases this is not true. For example,
it was found that oxygen exhibits a peak in the muon 
transfer rate $\Lambda_{\mu p \rightarrow \mu O}$ (in the following we use the abbreviate form ``$\Lambda_{pO}$'') at the epithermal energy~\cite{werthmuller96}.

The transfer to a heavier atom, $\mu^- p + ~^{A}\!Z \rightarrow$ $
({\mu^-} ~^A\!Z)^* + p$, leaves the muon into high orbital 
states of the newly produced muonic atom $({\mu^-} ~^A\!Z)^*$. The
muon atomic orbit equivalent to the electron $K$ orbit has a 
principal quantum number $n_\mu \approx (m_\mu/m_e)^{(1/2)} \approx 14$
and it is reached by the muon in a few femtoseconds from the
instant of its atomic capture \cite{mukhopadhyay}. The muon cascades down rapidly to the lowest
quantum state available, $1S$. In the case of light elements,
de-excitation starts with Auger process until the quantum number
reaches values from 3 to 6 at which radiative
transitions take over. During radiative transitions, X-rays are emitted at an energy corresponding to
the energy levels difference. In the oxygen case, these radiative
transitions have energies of the order of $\approx$100~keV and are easily detectable.

Thus, by adding small quantities of oxygen to hydrogen, one can observe the number of hyperfine transitions, which take place from the 
muon-transfer events, by measuring the time distribution of the 
characteristic X-rays of the added gas.

In order to fully exploit this effect, the energy dependence of the
muon transfer rate from muonic hydrogen to oxygen must be precisely
determined. This measurement was performed as a preparatory step of
the FAMU experiment. Details of the data analysis concerning the muons transfer rate
measurement are reported in this paper.

\section{Apparatus setup}
The setup of this FAMU data taking, performed in 2016, is described in details
in \cite{bonesini18}. A beam hodoscope was placed in front
of a high pressure gas contained in a thermalized aluminium vessel, surrounded by X-rays detectors. 
\begin{figure}[!tbh]
\begin{center}
\includegraphics[width=0.41\textwidth]{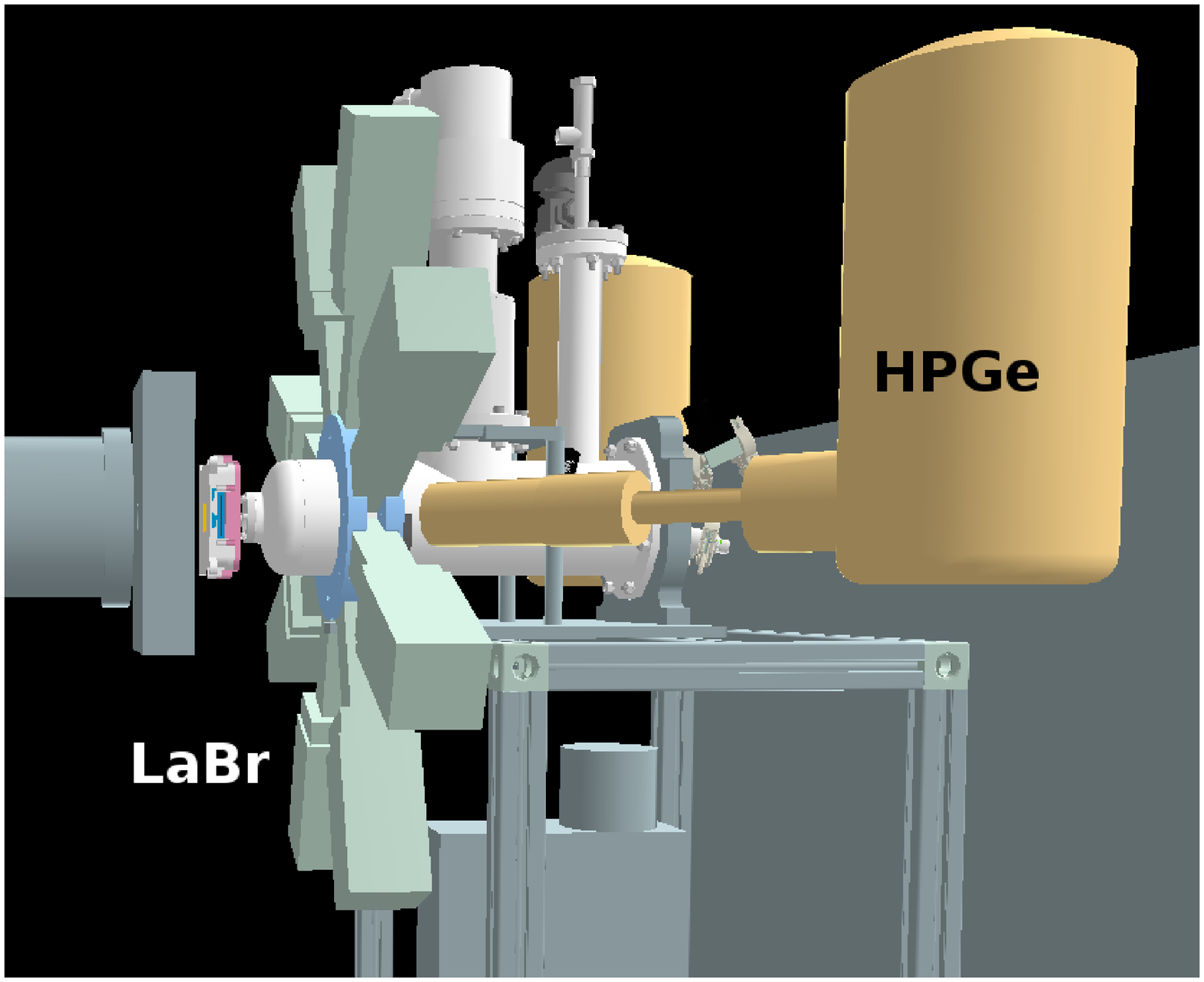}
\includegraphics[width=0.47\textwidth]{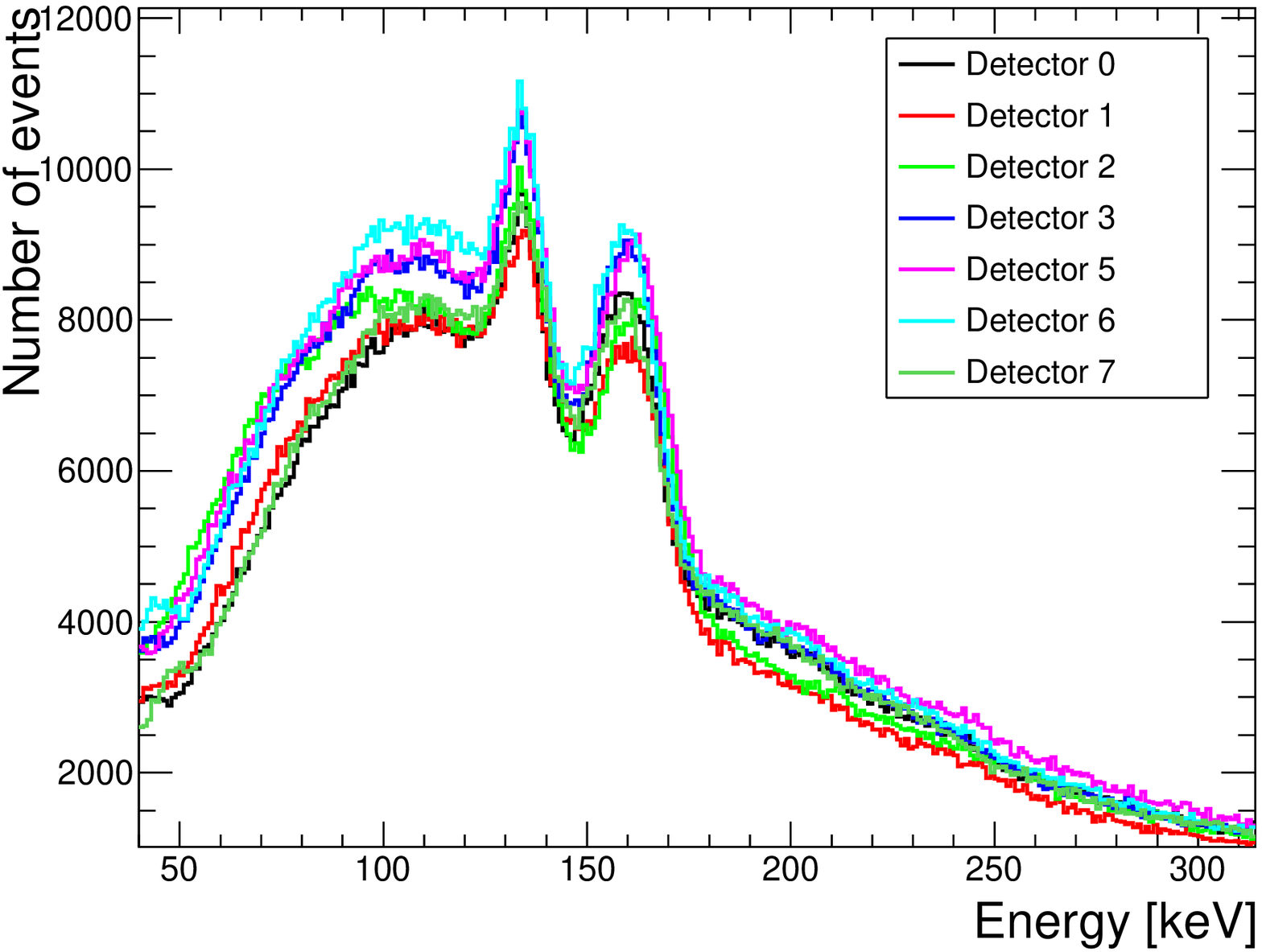}
\end{center}
\caption{\label{fig:CADcalib}Left panel: CAD drawing of the FAMU experimental setup. The
  star with eight LaBr$_3$(Ce) detectors is shown together with two
  HPGes. Muons entered the target from the left side of the
  figure. Right panel: energy spectrum of each LaBr$_3$(Ce) detector
  for delayed events. The $K_{\alpha} (133$~keV$)$ and the $K_{\beta}
  (158$~keV$) + K_{\gamma} (167$~keV$)$ lines can be observed.}
\end{figure}
Figure~\ref{fig:CADcalib}, left panel, shows a CAD drawing of the
experimental setup. Muons enter the
target from the left side (where the beam pipe tube is shown). 
The FAMU experimental method requires a detection system suited 
for time resolved X-ray spectroscopy \cite{adamczak16}. The characteristic X-rays from muonic atoms formed in 
different targets were detected using HPGe detectors (orange in
the drawing) and eight 
scintillating counters based on LaBr$_3$(Ce) crystals (star-shaped
structure, almond green in the drawing), whose outputs were 
recorded for 5~$\mu$s using a 500 MHz digitizer to measure both energy and 
time spectrum of the recorded events. With a detailed pulse analysis, the expected characteristic X-rays 
and lifetimes of various elements present in the setup were measured. 
For this measurement the target was filled with a mixture of hydrogen
and oxygen $H_2+0.3\%O_2$, measured by weight. The hydrogen gas contained a natural
admixture of deuterium (135.8$\pm$0.1 ppm~\cite{boschi}). The target was thermalized using a Sumitomo
helium cold head. In this case six temperatures were kept stable for
three hours each, from 100 to 300~K degrees.

In this analysis, the characteristic X-rays 
of muonic atoms were detected using scintillating counters based on 
LaBr$_3$(Ce) crystals (energy resolution $\approx$3\%
at 662 keV and decay time $\tau=16$~ns) read out by Hamamatsu
R11265-200 PMTs. The acquired waveforms were processed 
off-line to reconstruct the time and energy of each detected X-ray.

Data taking was performed at the Riken muon facility of the Rutherford Appleton Laboratory (RAL), in
Harwell (UK). Muons are produced in bunches with a repetition rate of 50~Hz. Each
bunch is consisting of two spills separated by about 320~ns. Each spill can
be roughly described as a gaussian distribution with FWHM of 70~ns
and the starting muon momentum has a gaussian distribution with
$\sigma_E/E \approx 10\%$~\cite{RAL}.
Beam users can tune the mean of the momentum distribution, which was chosen to be 57~MeV/c for the 2016 FAMU data taking.

Exiting the beam pipe collimator kapton window at a rate of about $10^5$ particle per second,
muons crossed the FAMU beam hodoscope~\cite{bonesini17} and reached the aluminium vessel. They
crossed a first external aluminium window, 0.8 mm thick, of the
cryogenic container before entering the aluminium target window,
2.8~mm thick, which contained the gas. The gas mixture was prepared at
41~bar at 300~K. The thermal cycle was performed keeping a constant density.

\section{Experimental method}
The transfer rate measurement as function of the muonic hydrogen
kinetic energy must be performed in a thermalized condition. In this
state, the kinetic energy of the $\mu p$ follows a Maxwell-Boltzmann
distribution at a given target temperature. 

However, soon after the muon capture, the $\mu p$ atom gains kinetic
energy due to its de-excitation. Data were taken with a gas density
such as the thermalization of muonic hydrogen, following the muon
capture, requires less than about 150~ns~\cite{bakalov15}.  The 0.3\% concentration (by weight) of $O_2$  used in the gas
mixture, was chosen in order to have a mean transfer time greater than
150~ns. This allowed the system to be fully thermalized after few
hundred nanoseconds from the arrival of the last muon. Under these
conditions, the average kinetic energy of muonic hydrogen can be
assumed as the Maxwellians corresponding to the gas temperature.

In the 2016 experimental setup, differently to the previous data
set~\cite{adamczak16, mocchiutti18}, a 
persistent X-ray emission from muonic oxygen de-excitation was detected
in the delayed phase, from 300 to about 5000~ns after the muon spill.

Hence, a study of the time evolution of the oxygen line emission can be
performed to measure the transfer rate. Since the transfer process
occurred in the delayed phase, in this analysis the large
background of X-rays prompt emission from all the
elements of the target - mostly aluminium, nickel, gold, and carbon -
was strongly suppressed and it was negligible.

The variation of number of muonic hydrogen atoms $N_{\mu p}$ present in the target in the time interval $dt$ can be
expressed by:
\begin{equation}
  dN_{\mu p}(t) = S(t) dt - N_{\mu p}(t)\lambda_{dis} dt,
\label{eq:tevol}
\end{equation}
where $S(t)$ is the number of muonic hydrogen generated in the
time interval $dt$, and $\lambda_{dis}$ is the total disappearance
rate of the muonic hydrogen atoms:
\begin{equation}
\lambda_{dis} = \lambda_0+\phi\,(c_p\Lambda_{pp\mu}+  c_d\Lambda_{pd} +
c_{O}\Lambda_{pO}).
\label{eq:ldis}
\end{equation}
Here $\lambda_0$ is the rate of disappearance of the muons
bounded to proton (that includes both muon decay and nuclear
capture), $\Lambda_{pp\mu}$ is the formation rate of the $pp\mu$ molecular ion in
collision of $\mu p$ with a hydrogen nucleus, $\Lambda_{pd}$ denotes the
muon transfer rate from $\mu p$ to deuterium, 
 and $\Lambda_{pO}$ is the muon transfer
 rates from $\mu p$ to oxygen.
 The $pp\mu$ formation and muon transfer rates
 are all normalized to the liquid hydrogen number density (LHD)
 $N_0=4.25\times 10^{22}$~cm$^{-3}$, and $\phi$ is the target gas
 number density in LHD units.
 The atomic concentrations of hydrogen, deuterium, and
oxygen in the gas target, indicated by
 $c_p$, $c_d$, $c_{O}$,
are related to the number
densities of the latter, $N_p$, $N_d$, and $N_{O}$, by:
\[ c_p=N_p/N_{tot}\ ,\ c_d=N_d/N_{tot}\ ,\ c_{O}=N_{O}/N_{tot}\ , \] 
\[N_{tot}=N_p+N_d+N_{O}\ ,\  c_p+c_d+c_{O}=1\ . \]
 The unknowns in Eq.~\ref{eq:tevol} and \ref{eq:ldis} are therefore $S(t)$ and
 $\Lambda_{pO}$. In this analysis only the delayed phase is considered
 so, by definition of delayed phase, $S(t)$ is a constant term and
 from Eq.~\ref{eq:tevol} it is possible to extract the transfer rate $\Lambda_{pO}$.
Gas parameters depend on the composition, temperature, and pressure as
described below; values were set as follows: 
\begin{itemize}
\item filling was performed at $T=300$~K,
 at a pressure $P=(41.00\pm0.25)$ bar, with $O_2$ concentration of
  $(0.30\pm0.09)$\%, mass weighted;
\item the number densities of the gas mixtures are
  $\phi = (4.869\pm0.003)\times10^{-2}$ in LHD atomic units, as derived from
  previous values; 
\item the atomic concentrations of oxygen $c_{O} = (1.90\pm0.04)\times 10^{-4}$ is derived from previous values; 
  the deuteron concentrations $c_{d}=(1.358\pm0.001)\times 10^{-4}$ was obtained from a laboratory measurement \cite{boschi};
\item remaining data
  were taken from literature and theoretical calculations:
  $\Lambda_{pp\mu} = 2.01 \times 10^6$ s$^{-1}$~\cite{andreev15},
  $\Lambda_{pd} = 1.64 \times 10^{10}$ s$^{-1}$ as function
  of temperature~\cite{adam17}, $\lambda_0 =(4665.01\pm0.14)\times 10^2$~s$^{-1}$~\cite{andreev15,suzuki}.
\end{itemize}

A fit of the oxygen X-rays time evolution can be performed
numerically integrating Eq.~\ref{eq:tevol} by leaving $\Lambda_{pO}$
as free parameter. 

\section{Data analysis}
The experimental sample, used in 
this work, consists of about $2.6\times 10^6$~muon triggers (double spills),
corresponding to $\approx 7.8\times 10^7$ reconstructed X-rays coming
from seven out of eight LaBr$_3$(Ce) detectors. One LaBr$_3$(Ce) detector output was
not used in this analysis due to hardware problems. Target temperature was set at 100, 150,
200, 240, 273 and 300~K for three hours. A precise temperature measurement
(relative error $\sigma_T/T$ better than $10^{-5}$) was performed using two probes and
it was recorded together with the data at each trigger. 
Eleven temperature bins were chosen, six corresponding to the set
temperatures and five corresponding to intermediate values with
rapidly changing temperature. The time needed to the system to
stabilize at a set temperature varied from half hour to about an hour
and a half, depending on the set temperature and on external conditions.

To extract the transfer rate as function of the kinetic energy of the
$\mu p$ atom, the following procedure was used:
\begin{enumerate}[label=\arabic*)]
\item LaBr$_3$(Ce) detectors were calibrated and their signals were integrated all together.
\item Good events were selected from the data set and selection
  efficiencies were evaluated.
\item The delayed events of each data set corresponding to a temperature bin were recorded into
  several time bins. For each time bin the energy spectrum in the range from 50 to
  500 keV was obtained, in fact we considered the $K_{\alpha} (133$~keV$), K_{\beta} (158$~keV$),
K_{\gamma} (167$~keV$)$ X-ray lines from oxygen when evaluating
$\Lambda_{pO}$. For each energy spectrum, a background was evaluated and
  subtracted.
\item For each temperature the set of data was fitted, as explained in
  the previous section, obtaining a value for the transfer rate.
\item For every set of data corresponding at a given temperature bin,
  the mean kinetic energy of the $\mu p$ was evaluated.
\end{enumerate} 
Whenever needed, possible sources of systematical errors were investigated. Each item of this procedure is explained with more details in the following.

\subsection{Detector calibrations}
Signals coming from the LaBr$_3$(Ce) detectors were saved as waveforms by
the acquisition system. A C++/ROOT based software was used to
distinguish the pulses and to reconstruct their energy by fitting the
waveform~\cite{bonesini18}. To each detected X-ray signal a time start and an energy in
ADC channel was associated. The calibration from ADC channels to energy
was performed independently for each detector using the data set
itself.

Prompt and delayed X-ray lines were used first to correct for any
fluctuation due to temperature variation inside the experimental
hall. These were small long term variations of few keV in several
hours due to the read-out electronics. This correction was needed to improve the overall energy
resolution and to normalize the response of the behaviour of different
detectors.

The ADC spectrum for each detector was converted to energy
by comparing the position of the detected peaks to the known X-ray
lines. We used a $\chi^2$ minimization algorithm on the energy region
of interest. Calibration results are shown in Fig.~\ref{fig:CADcalib},
right panel, for the seven
detectors used in this analysis. Each line color corresponds to a
different detector. The energy spectra are not normalized. The different
number of events detected depended on the detector position around the
target. In fact, the overall cylindrical symmetry of the system was broken by the
position of the insulating materials and of the copper braids used to
thermalized the inner pressurized vessel. From the figure it can be
noticed that the $K_{\beta}$ and $K_{\gamma}$ lines cannot be
resolved, being the energy resolution of these detectors about 9\%~(FWHM)~\cite{bonesini18}.

In this analysis, the energy spectra of the seven
detectors were summed to increase the statistics.

\subsection{Data selection and selection efficiencies}
Each trigger corresponded to an acquisition window of about
10~$\mu$s. Since the data analysis consists in the study of the time
evolution of oxygen X-rays signals in this time window, each
time-dependent effect was carefully studied and taken into account. In
this experimental setup the delayed phase started at 1200~ns, about
300~ns after the arrival of the second muon pulse. We refer to events from time
zero to 1200~ns as to ``prompt'' emission.

The reconstruction software read the waveform, recorded every 2~ns, in
order to detect and process every single pulse. Pulse recognition was
performed by setting a fixed threshold on the derivative of the
waveform. The software was able to detect with very high efficiency
X-rays with energy greater than few keV and separation greater than
the signal rise time (12~ns). A realistic simulation of the waveform showed that
the software recognized X-rays with energy greater than 20~keV and
separation greater than 15~ns with an efficiency greater than 99.9\%.

Once defined the start time of each pulse, the software performed a fit
of the waveform in order to measure correctly the amplitude
(i.e. the energy) of piled up signals. The number of piled up signals
varied in the trigger window.  
\begin{figure}[!htb]
\begin{center}
\includegraphics[width=0.48\textwidth]{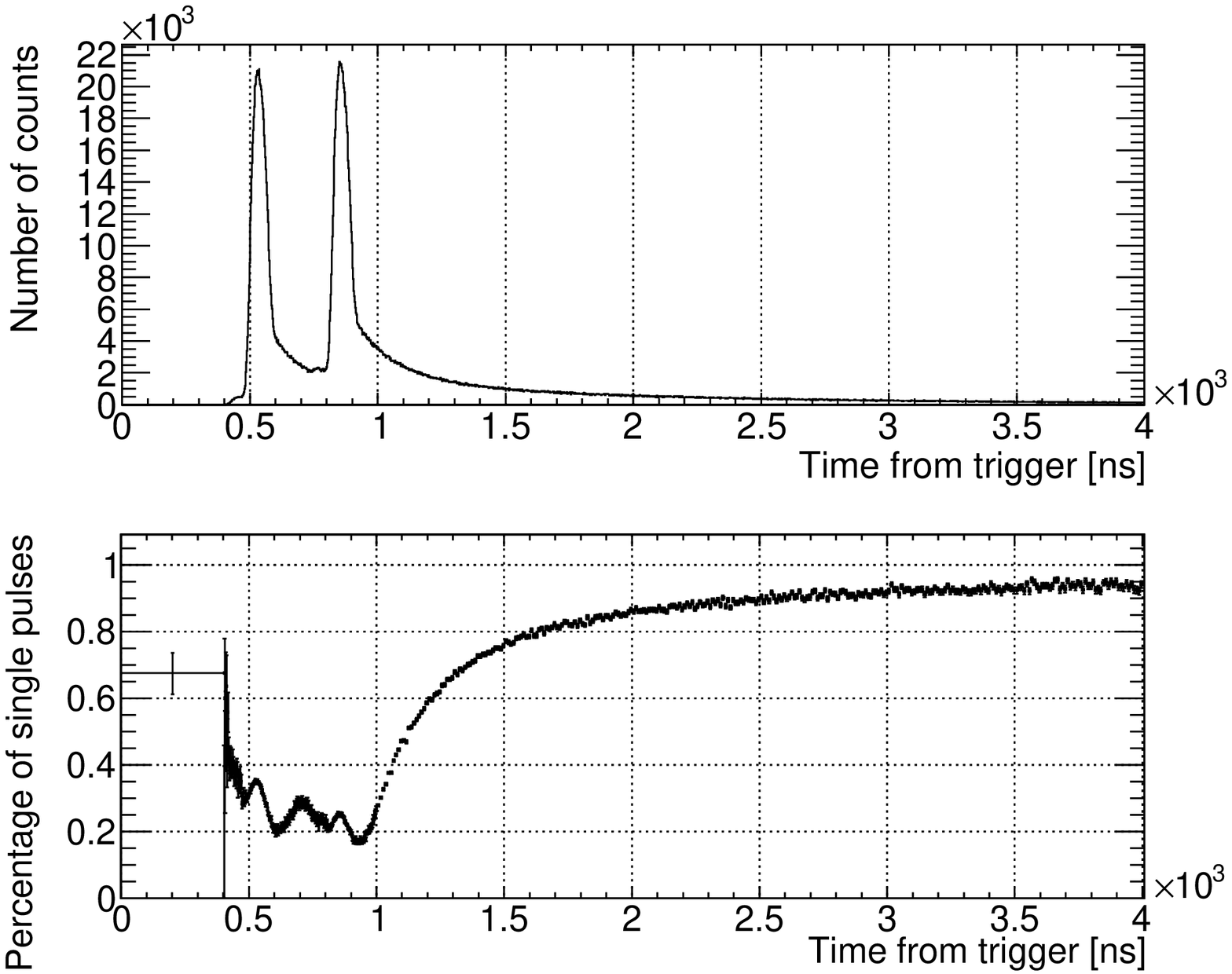}
\includegraphics[width=0.48\textwidth]{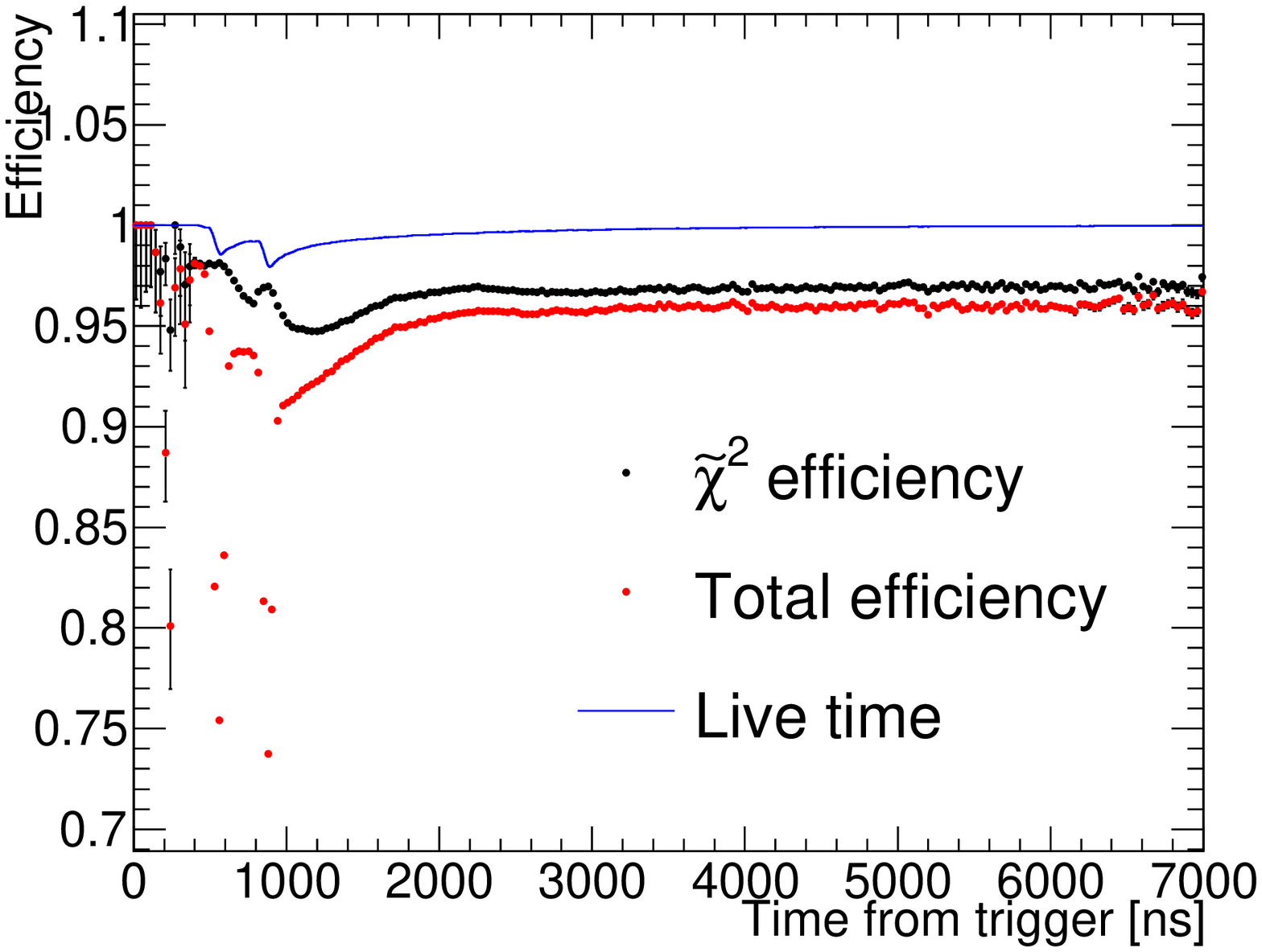}
\end{center}
\caption{\label{fig:effi} Top left panel: time distribution of X-rays
  reconstructed events. Bottom left panel: percentage of single pulses as
  function of time. Right panel: selection efficiencies (circles)
  and detector fractional live time (blue line).}
\end{figure}
Figure~\ref{fig:effi}, panel on the top left, shows the counts of
detected X-rays as function of the time. The two peaks correspond to
the muon arrival time and they have the same profile of the muon beam,
i.e. 70~ns FWHM and separation of 320~ns. The percentage of single
pulses, as detected by the reconstruction software, is shown in the
bottom left panel. In the delayed phase, the number of piled up events
becomes soon very small. Consequently the probability of having
piled up events with separation smaller than 12~ns (``perfect'' pile
up) was considered negligible. 

Weak selection were applied on the reduced $\chi^2$ ($\tilde{\chi}^2$) coming from the
fitting algorithm and on the signal separation. These selections were
chosen in order to reject badly reconstructed events keeping as flat as possible the selection efficiency as
function of time. Since the correlation between the start time and the
fit result is weak, it was possible to estimate the selection
efficiency using the data set itself - at least in the delayed phase
where the reconstruction efficiency is close to the unity. 
Figure~\ref{fig:effi}, on the right, shows the resulting $\tilde{\chi}^2<100$ selection
efficiency as function of the time (black circles). The separation selection efficiency (distance between signals greater than
30~ns) was evaluated on the events passing the $\tilde{\chi}^2$ selection
and it was close to 99\% efficiency (not shown in figure). The overall efficiency was of about 96\% and it was almost time
independent in the delayed phase, red circles in figure~\ref{fig:effi}
(left panel).

The last possible time dependent effect was the live time of the
detectors. LaBr$_3$(Ce) crystals and their electronics are very fast,
however in this experiment their energy range was limited to about 800
keV by the number of channels of the digitizer. Signals with higher energy deposit
saturate the ADC channels, giving a plateau on the
waveform. The width of the plateau depended on the energy
release. During this time the detector was not able to detect other
signals, hence we defined these periods as ``dead time''. This
dead time varied in the trigger window depending on the pile-up and on
the type of particles hitting the detector (charged particles usually
give a saturated signal). The live time and dead time was measured
using the data
themselves and an overall average correction was applied. Live time
was measured for each detector independently and the live time correction was
applied before summing the detectors signals. The LaBr$_3$(Ce) detector
were chosen for their fast response, consequently the live time is
very high. An example of fractional live time for one detector and a subset of
data is shown in Fig.~\ref{fig:effi}, right panel blue
line. Correction on the data was always less than 2\%.

\subsection{Background estimation} 
Spectra of the selected data sample were corrected for all the
found efficiencies and for the fractional live time. Then, the data
were sampled into eleven temperature bins.

For each temperature bin the energy spectrum of delayed events was
studied as function of the time. The delayed time window from 1200 to
10000~ns was split into 20 bins. 
\begin{figure}[!htb]
\begin{center}
\includegraphics[width=0.48\textwidth]{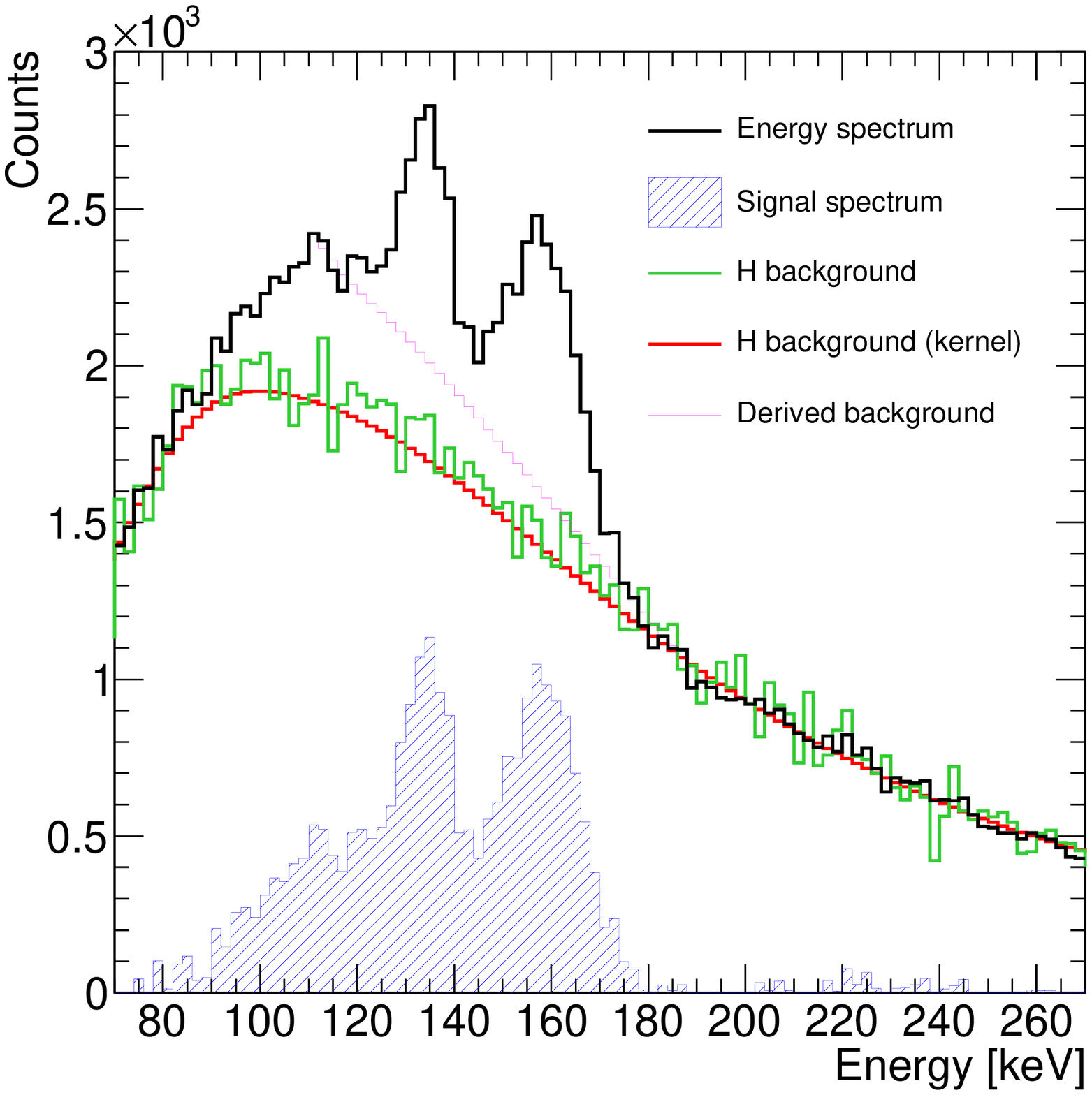}
\includegraphics[width=0.48\textwidth]{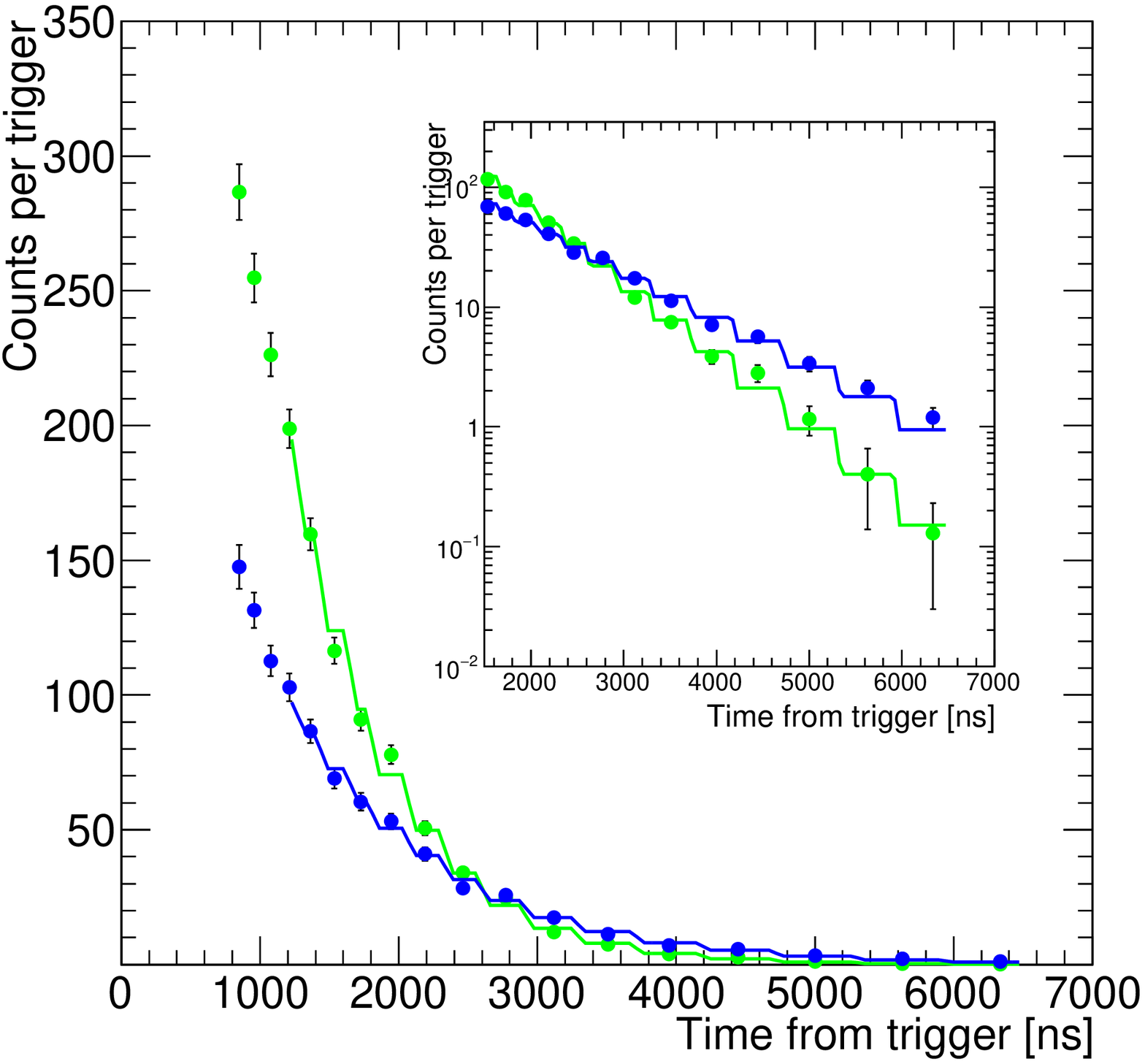}
\end{center}
\caption{\label{fig:EspFit}Left panel: example of energy spectrum
  signal in the time bin [1450,1650]~ns at 300~K. Right panel: time
  dependence of the oxygen signal at 300~K (green points and line) and
  at 100~K (blue points and line). Lines are fit to the
  data. Error bars represent statistical and background-related
  systematics summed quadratically. Inset: zoom of the figure in logaritmic scale.}
\end{figure}
An example of a delayed energy spectrum is shown in
Fig.~\ref{fig:EspFit}, black line in the left panel, at the
temperature of 300~K and the time bin from 1450 to 1650~ns. The oxygen
lines are seen over the background. The background estimation and
subtraction for each time and temperature bin was the most subtle
point of this work. Detector signals, in fact, showed a tail towards
low energies due both to physics (Compton energy losses in the
supporting material) and to electronics (most probably small gain
variations due to the muon beam rate). Unfortunately, in the case of
the oxygen mixture, the
tail goes down to about 100 keV, where the energy spectrum naturally bend
towards lower values. In such a situation it was impossible to estimate
the background directly from the data. The background derived as an interpolation, or even more
complex spectroscopic algorithms like the one provided by the ROOT
class ``TSpectrum''~\cite{morhac}, gave an overestimated background
(e.g. magenta line in Fig.~\ref{fig:EspFit}, left panel). Better
results were obtained using the spectrum detected with the target
filled with pure hydrogen, at the same temperature and pressure. The
hydrogen background is shown as a green line in figure. However, the
acquired statistics with pure hydrogen was about one tenth of the gas
mixture one. The low statistics in the background subtraction resulted in higher
fluctuations on the signal data. In order to reduce these fluctuations
the hydrogen data set was smoothed using a gaussian kernel
algorithm~\cite{hastie}, with smoothing parameter set to the energy
resolution of the LaBr$_3$(Ce) detectors. The result is shown as a red
line in Fig.~\ref{fig:EspFit}, on the left, and it was used to get the signal
spectrum (hatched blue line histogram). 

Fluctuations due to the normalization of the background and to the type of background subtraction (raw or
with kernel smoothing method) were considered as systematic
errors. This type of systematics was associated to each spectrum,
i.e. each time bin of each temperature. The net result was a systematical 
fluctuation of the measured transfer rate around the true value, with
no absolute trend. We compared backgrounds obtained with different
methods and parameters and the results obtained with
different normalization regions for the background. These tests
permitted us to estimate the effect of this systematic between 5\% and
20\% on the integrated signal spectrum (depending on the bin
statistics). We decided to sum quadratically this systematic with the
statistical error for each signal spectrum, before performing the fit to the
data that provides the transfer rate measurement.

\subsection{Data fit and systematical uncertainties evaluation}
Time dependence for oxygen lines at the two temperature extremities --
300~K (green points) and at 100~K (blue points) -- is shown in
Fig.~\ref{fig:EspFit} (right panel). Each 
point represents the integrated signal after the background
subtraction for each time bin, divided by the bin width. The fit was performed between 1200~ns, when the delayed
phase started, and 6500~ns, when the signal was still detectable above
the background. The fit to the data are shown as solid lines. The step-like behaviour is due to the numerical
integration procedure used to calculate the function for each time
bin. It can be noticed how the slope of the two distribution is
significantly different, corresponding to different transfer rate
measurements as function of the temperature. 

As explained previously, the error bars associated to the points include both statistical and
background-systematic errors summed quadratically. 

Since the fit of the data was performed by numerically integrating
Eq.~\ref{eq:tevol}, uncertainties on the equation parameters result in
additional systematical uncertainties on the measured transfer rate. 
The estimation of systematical effects was done by varying the
parameter values by the known errors. Systematic uncertainties were considered negligible whenever smaller
than 1\%. 

The gas used to create the mixture were high purity gases, 99.9995\%
pure, corresponding to a contamination of other gases smaller than
5~ppm. The simulation proved that such a 
contamination imply negligible differences in the results.
However, the gas mixture was prepared by the gas supplier by weight
with a relative error of 3\%. Propagation of this uncertainty in the
fitting formula brings both different atomic concentrations and gas
density. The overall effect on the final measurement is of the order
of about 3\%.

The gas filling procedure for this mixture was erroneously performed
without a precise pressure reading. Consequently the uncertainty on
the gas density had an effect on the transfer rate measurement of the
order of about 3\%.

All other items considered as possible source of systematics were proved to
have negligible effects to the final measurement. These included the
error (given in the literature) in the determination of the muon decay time
(proton-bounded) $\lambda_0$; the literature error in the
determination of transfer rate to $pp\mu$ molecule $\lambda_{pp\mu}$;
the error in the determination of transfer rate to deuterium
$\lambda_{pd}$; 
the error in the isotopic composition of hydrogen gas $c_d$.

Other consistency check were also performed on data by selecting
separately $K_\alpha$ and $K_{\beta,\gamma}$ lines, by varying time
fit window, and by applying much stricter selection criteria
(e.g. $\tilde{\chi}^2<10$ and pulse time separation greater than
100~ns). All these checks resulted in a smaller data set and in a
consequent statistical fluctuation around the higher statistics value.

\subsection{Temperature to kinetic energy conversion}
Given a fixed temperature, the conversion to kinetic energy can be
performed by calculating the mean of a Maxwell--Boltzmann
distribution. In our data set, however, the temperature slowly varied
around a set value. The variation was due to the thermal capacitance
of the target. To reduce the acquisition time and increase the
statistics we decided to use the whole data set instead of waiting for
a thermalization better than a given temperature threshold. Hence,
for each trigger the target temperature was recorded and was
associated to the detected X-rays. 

Therefore, the measurement of the mean kinetic energy was obtained by
the following procedure. For each X-ray passing
the data selection and with energy inside the integrating region
(both signal and background events) a Maxwellian function was
calculated. The Maxwellian functions of all the triggers recorded at a
given temperature bin were summed. The mean of the resulting
distribution was chosen as a measurement of the kinetic energy. Final
results will be presented as function of the measured kinetic energy.

\section{Results}
Preliminary transfer rate results are shown, as function of the
temperature, 
\begin{figure}[!htb]
\begin{center}
\includegraphics[width=0.8\textwidth]{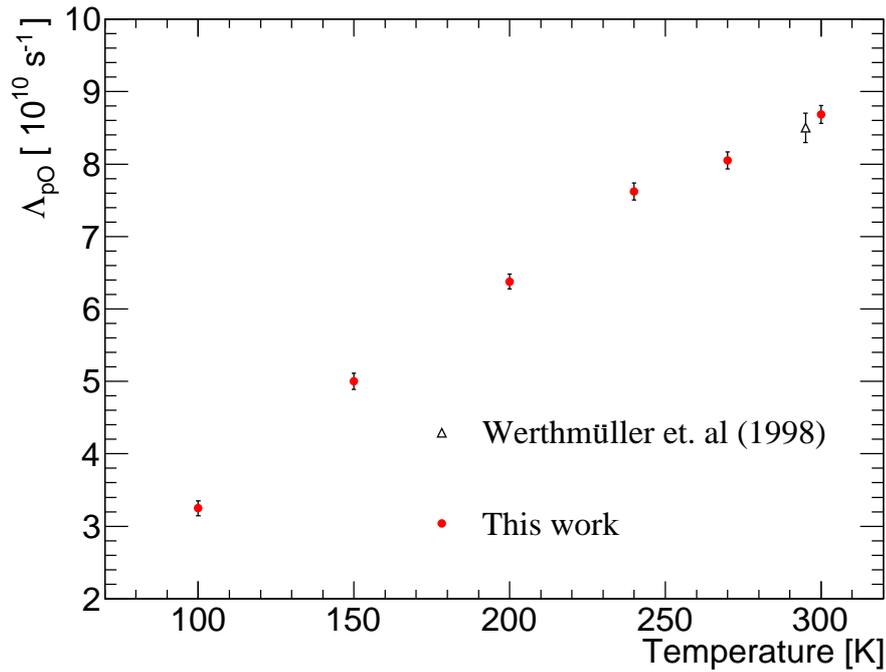}
\end{center}
\caption{\label{fig:results}Transfer rate measurement as function of
  the temperature. The result obtained by Werthm\"uller et al. at PSI at 294~K~\cite{wertmuller}
  is also shown.}
\end{figure}
in Fig.~\ref{fig:results}. This is the first precise measurement of the
temperature dependence of transfer rate from muonic hydrogen to
oxygen. Our results are in excellent agreement with the previous
measurement at 294~K~\cite{wertmuller}. Error bars represent the error
on the fit result obtained as described in previous section, i.e. errors include statistical and background-related
systematics. Other systematics are not included.

\section{Conclusions}
The experimental results described here represent a fundamental step
in the development of the FAMU project.

Data were taken in 2016 with a dedicated cryogenic thermalized gas
target and with optimal concentrations. The confirmation of the
theoretically predicted temperature dependence of the transfer rate
from muonic hydrogen to oxygen will permit to proceed with the measurement
of the muonic hydrogen ground state hyperfine splitting. Moreover, this
precise measurement will allow to optimize the target and detection
system for the final experiment.

Other elements gas mixtures data taken in the same run are under
study and results will be the topic of future publications.

\section*{Acknowledgments}
The research activity presented in this paper has been carried out in the framework of the
FAMU experiment funded by Istituto Nazionale di Fisica Nucleare (INFN). The use of the
low energy muons beam has been allowed by the RIKEN RAL Muon Facility. We thank the RAL
staff (cooling, gas, and radioactive sources
sections) and especially Mr. Chris Goodway, Pressure
and Furnace Section Leader, for their help, suggestions,
professionalism and precious collaboration in the set up of the experiment at RIKEN-RAL.

We gratefully recognize the help of T. Schneider, CERN EP division,
 for his help in the optical cutting of the scintillating fibers of the
 hodoscope detector and linked issues and N. Serra from
 Advansid srl for useful discussions on SiPM problematics.

We thank our colleagues Chiara Boschi and Ilaria Ba\-ne\-schi (IGG, CNR Pisa) for their help in
the measurement of the gas isotopic composition.

A. Adamczak and D. Bakalov acknowledge the support within the bilateral agreement
between the Bulgarian Academy of Sciences and the Polish Academy of Sciences. D. Bakalov,
P. Danev and M. Stoilov acknowledge the support of Grant 08-17 of the Bulgarian Science
Fund.



\end{document}